# Small Cell Deployments: Recent Advances and Research Challenges

Zubin Bharucha, DOCOMO Eurolabs, Germany,
Emilio Calvanese, CEA-LETI, France,
Jiming Chen, Ranplan Wireless Network Design Ltd, UK
Xiaoli Chu, University of Sheffield, UK,
Afef Feki, Bell-Labs, France,
Antonio De Domenico, CEA-LETI, France,
Ana Galindo-Serrano, CTTC, Spain,
Weisi Guo, University of Sheffield, UK,
Raymond Kwan, Ubiquisys, UK,
Jimin Liu, Ranplan Wireless Network Design Ltd, UK,
David López-Pérez, King's College London, UK,
Massod Maqbool, SAGEMCOM, France,
Ying Peng, Datang Telecom Group, China,
Samir Perlaza, Princeton University, USA,
Guillaume de la Roche, Mindspeed Technologies, France,
Serkan Uygungelen, DOCOMO Eurolabs, Germany,
Alvaro Valcarce, Triagnosys, Germany,
and Jie Zhang, University of Sheffield, UK

*Abstract*—**This paper summarizes the outcomes of the 5th International Workshop on Femtocells held at King's College London, UK, on the 13th and 14th of February, 2012. The workshop hosted cutting-edge presentations about the latest advances and research challenges in small cell roll-outs and heterogeneous cellular networks. This paper provides some cutting edge information on the developments of Self-Organizing Networks (SON) for small cell deployments, as well as related standardization supports on issues such as carrier aggregation (CA), Multiple-Input-Multiple-Output (MIMO) techniques, and enhanced Inter-Cell Interference Coordination (eICIC), etc. Furthermore, some recent efforts on issues such as energy-saving as well as Machine Learning (ML) techniques on resource allocation and multi-cell cooperation are described. Finally, current developments on simulation tools and small cell deployment scenarios are presented. These topics collectively represent the current trends in small cell deployments.**

## I. INTRODUCTION

DUE to the proliferation of smart mobile devices and innovative mobile data services, telecommunication systems are facing increasing demands for ubiquitous heterogeneous broadband mobile communications. It has been predicted that the volume of wireless data will exceed that of wired data by 2015. In order to realize the enormous data capacity and meet user Quality of Service (QoS) requirements, while keeping operators' costs low, low-power low-cost small cells operating in licensed spectrum, have been widely considered as the most promising solution. Small cells include femtocells, picocells, metrocells and microcells, in order to increase the cell coverage and capacity. Due to their low cost and easy deployment, small cells provide a viable and cost-effective way to improve cellular coverage, capacity and applications for homes, enterprises, as well as for metropolitan and rural areas.

Driven by their attractive features and potential advantages, the development and deployment of small cells have gained tremendous momentum in the wireless industry and research communities in recent years. Small cells have also attracted the attention of standardization bodies, e.g., the 3rd Generation Partnership Project (3GPP), Long Term Evolution (LTE) and LTE-Advanced. However, it is worth noting that the successful rollout and operation of small cells are still facing significant technical challenges and issues.

The aim of this paper is to summarize the outcomes and discussions of the 5th International Workshop on Femtocells held at King's College London, UK, on the 13th and 14th of February, 2012. The workshop hosted cutting-edge presentations about the latest advances and research challenges in small cell roll-outs and heterogeneous cellular networks.

The rest of the paper is organized as follows. In Section II, the standardization activities regarding small cells within 3GPP are reviewed. In Section III, the self-organization features and related techniques of small cells are discussed. Section IV presents 3GPP LTE-Advanced small cell Carrier Aggregation (CA) techniques and performance evaluation. In Section V, enhanced Inter-Cell Interference Coordination (eICIC) is proposed for small cells and the related performance is evaluated through simulations. In Section VI, Multiple-Input-Multiple-Output (MIMO) techniques are explored for small cells to achieve spatial multiplexing. In Section VII, the trade-off between spectral, energy and cost efficiencies offered by the new architecture of small cell networks is studied. Section VIII reviews Radio Resource Management (RRM) techniques based on Reinforcement Learning (RL) techniques for small cells. Section IX introduces techniques enabling decentralized operation of small cells. In Section X, a comprehensive LTE system-level simulator is presented. In Section XI, static and mobile deployment scenarios of small cells are discussed. Conclusions are drawn in Section XII.

## II. 3GPP STANDARDIZATION

The International Telecommunications Union (ITU) officially issued the first release of the 4th Generation (4G) mobile telecommunications systems - International Mobile Telecommunications (IMT)-Advanced standard - in March 2011. Meeting the rapid increase in mobile data traffics is a crucial challenge for any future mobile systems. Therefore, ITU has raised harsh requirements on peak data rates and spectrum efficiency for IMT-Advanced systems [2], e.g., peak data rates above 600 Mbps, a Downlink (DL) peak spectrum efficiency of 15 bps/Hz, and an Uplink (UL) peak spectrum efficiency of 7.5 bps/Hz. In October 2009, the LTE-Advanced, evolution of the 3rd Generation (3G) Wideband Code Division Multiple Access (WCDMA) and Time-Division Synchronous Code Division Multiple Access (TD-SCDMA) systems, was submitted to the ITU as an IMT-Advanced candidate system and thereafter as a 4G standard system [1].

In order to meet IMT-Advanced standard requirements with limited radio resources, deploying Low Power Nodes (LPNs), such as Remote Radio Heads (RRHs), picocells, femtocells and relay nodes, overlaid on the existing macrocell network has been heralded as the most promising solution. Therefore, this type of networks, referred to as Heterogeneous Networks (HetNets), has been and will continue to be widely investigated in 3GPP. The HetNet concept has been included into a number of study/work items in LTE-Advanced, in conjunction with advanced technologies such as CA, non-CA based eICIC and its improvements, i.e., Coordinated Multi-Point transmission and reception (CoMP), and enhanced PDCCH (ePDCCH). Moreover, in LTE-Advanced, Evolved-Universal Terrestrial Radio Access Network (E-UTRAN) protocols and network interfaces are also being enhanced.

In LTE Rel-10 (the first version of LTE-Advanced), CA supports *cross carrier scheduling*, in which User Equipments (UEs) can detect PDCCH, PCFICH and PHICH from a subset of existing Component Carriers (CCs). This is an effective interference management technique, since it allows scheduling control channels in one CC while scheduling data channels in another CC. In this way, inter-cell interference to control channels can be efficiently mitigated. CA and cross carrier scheduling are also being investigated in LTE Rel-11 [3]. Whether or not inter Base Station (BS) signaling is required for robust and autonomous inter-cell interference management and how each BS selects those CCs that maximize the overall network performance are being studied.

*CA based* solutions are attractive for scenarios with large availability of spectrum and CA capable UEs. However, *non-CA* based solutions are still needed to enable HetNet operation in scenarios with limited spectrum and UEs without CA capability [4], e.g., co-channel deployments. Simple cell splitting and LTE Rel-8/9 ICIC in co-channel deployments with unbalanced transmit powers among BSs cannot provide sufficient cell-edge coverage and interference mitigation. Moreover, mechanisms allowing off-loading from macrocells to Low Power Nodes (LPNs), e.g., range expansion, may worsen cell-edge capacity. Hence, new interference management techniques are required for HetNet co-channel deployments. For example, the eICIC in LTE Rel-10 is mainly based on Time Domain (TD) resource partitioning through the use of Almost Blank Subframes (ABS), and introduces restricted RRM, Radio Link Monitoring (RLM) and Channel State Information (CSI) measurements to deal with varying inter-cell interference conditions in the time domain, while maintaining satisfactory DL and UL coverage. ABS patterns need to be carefully designed in order to protect control channels such as SCH/PBCH/SIB1/Paging and ensure Hybrid Automatic Repeat Request (HARQ) timing. ABS information can be exchanged between cells through the X2 interface. eICIC based on DL power control at femtocells and UL power control at UEs were discussed in LTE Rel-10 for inter-femtocell interference scenarios, but both power control approaches were considered as implementation issues.

Due to time limitation, some other identified eICIC techniques were de-prioritized in LTE Rel-10. Particularly, the impact of legacy transmissions on control and data channel demodulation when using ABS was left to be considered in LTE Rel-11, leading to Further enhanced ICIC (FeICIC). The ongoing work item on FeICIC attempts to identify

- UE performance requirements, possible air interface changes and enhanced Node B (eNB) signaling improvements, in order to enable significantly improved detection of Physical Cell Identity (PCI) and system information (e.g., MIB/ SIB1/Paging) in the presence of dominant interferers depending on UE receiver implementations;

- UE performance requirements and UE signaling required for both significantly improved DL control and data detection and UE measurement/reporting in the presence of dominant interferers depending on UE receiver implementations [5]; and

- network assistance information for Common Reference Signal (CRS) interference cancellation, improved detection of system information, and usage of reduced power ABS.

In LTE Rel-11, enhanced DL physical control channel(s) are also being investigated to allow control channel inter-cell interference mitigation.

CoMP was another technique proposed to enhance coverage, data rates, cell-edge throughput and system performance in LTE-Advanced. Its studies have been in progress since LTE Rel-9, with a work item set up in LTE Rel-11. CoMP scenarios 3 and 4 [6], for HetNets with low power RRHs overlaid on the macrocell network, extend the co-channel deployment ICIC research into data transmission enhancement. Instead of the semi-static coordination in eICIC via X2 interface, DL CoMP [6] implies dynamic coordination among multiple geographically separated transmitting and/or receiving points. DL CoMP techniques, such as simultaneous data transmissions from multiple points to a single or multiple UEs (i.e. Joint Transmission (JT)), Dynamic Transmission Point Selection (DTPS), Coordinated Scheduling (CS) and Coordinated Beamforming (CB), can turn interference into useful signals, and improve the received signal quality and/or data throughput. UL CoMP may facilitate multi-user transmissions, power control and detection of control channels in HetNets.

Besides inter-cell interference mitigation, 3GPP also works on protocol enhancement, UE mobility and network interfaces for HetNets. For example, in LTE Rel-9, the femtocell synchronization problem was widely studied. In LTE Rel-10, intra Closed Subscriber Group (CSG) femtocell mobility was enhanced, where X2-based handover between femtocells is allowed if no access control is required at the Mobility Management Entity (MME). In other words, handover is allowed between CSG or hybrid access femtocells with the same CSG identity, or when the target femtocell is an open access femtocell. Such handover assumes direct X2 connectivity between femtocells, no matter whether any of the involved femtocells is connected to a femtocell Gateway (GW) or not [7]. Seamless and robust mobility of UEs from macrocells to LPNs and vice versa is also considered in Rel-11 [9] to enable off-loading benefits.

Scenarios with multiple LPN-layers and carriers may require additional reestablishment procedures to improve overall system robustness. In this line, large-scale evaluation is needed for strategies to evaluate LPN discovery/identification and HetNet mobility performance under Rel-10 eICIC features. LTE Rel-11 is also evaluating the benefits of enhanced inter-CSG mobility, X2 connection via a GW for macrocell/femtocell to femtocell mobility enhancement, and deployment scenarios with two GWs directly interconnected to each other [8].

## III. SELF-ORGANIZATION

Self-organization has been considered as a key feature of LPNs. Since the number of LPNs to be deployed is expected to be large and LPNs are likely to be plug and play devices deployed by operators/users in an uncoordinated manner, it is not feasible for operators to optimize LPN deployments using traditional network planning and optimization techniques. Therefore, LPNs should adapt their parameters independently and autonomously on a regular basis depending on network, traffic and channel fluctuations. Self-organization has attracted much momentum in 3GPP standardization through a number of work items.

Self-organizing Network (SON) operations usually consist of three phases:

- Self-configuration, which is performed at start-up and configures initial parameters, e.g., a LPN initializes its transmit power according to the received power from the closest macrocell.

- Self-optimization, which is performed on a regular basis and tunes parameters according to network, traffic and channel fluctuations, e.g., LPNs may tune their handover thresholds depending on traffic load.

- Self-healing, which automatically detects network failures and corrects/mitigates them, e.g., if a cell is out of order, the neighboring cells may take over its traffic by enlarging their coverage.

Self-organization of a cell usually takes the parameters of neighbouring cells into account. For instance, self-organized inter-cell interference mitigation requires neighbouring cells to reduce coverage overlap while supporting seamless handovers. Self-organizing schemes can be generally categorized into the following two kinds:

- Centralized schemes, where a global entity is in charge of a given number of cells and is responsible for their optimization. In a centralized approach, LPNs can forward their configurations and measurements to a server, which will optimize relevant parameters based on overall network information.

- Distributed schemes, where each cell optimizes its own parameters based on local sensing and optimization techniques. Using distributed techniques, LPNs may learn about neighbouring cells and fine tune relevant parameters in a faster manner.

In practice, pure centralized schemes are difficult to implement, especially for a large number of cells. In such scenarios, the large number of parameters to be optimized and the associated significant signalling overhead may lead to high complexity. Distributed schemes face challenges arising from network, traffic and channel uncertainties, and inaccuracies in data obtained from neighbouring cells. Hybrid approaches, where some of the parameters are optimized in a centralized manner whereas others are optimized in a distributed manner, can be used to achieve a good trade-off.

Sensing plays a key role in the proper operation of distributed schemes, and can be performed in different ways:

- Network monitor mode: LPNs are able to scan, listen and estimate parameters from neighbouring cells.

- Measurement reports: it is proposed in LTE that UEs send measurement reports to their serving cells. When many UEs are located in the LPN coverage, the large amount of measurement reports collected can provide an accurate knowledge of its radio environment.

- Cooperative BSs: LPNs exchange information via a dedicated air interface or backhaul.

Each time the central server (in centralized SON) or the cell (in distributed SON) gathers fresh data, it can perform a new optimization if necessary. In 3GPP LTE-Advanced, the following use cases/parameters may be optimized:

- Coverage and capacity optimization: A typical SON task is to maximize network coverage and capacity. Cells size are typically optimized so that the overlap between neighbouring cells is minimized. Most existing solutions are related to transmit power optimization, based on the network monitoring mode, where a cell adapts its transmit power depending on the power received from the neighbouring cells so that its connected UEs can always maintain a useful Signal to Noise Ratio (SNR) level within a targeted radius.

- Energy saving: Energy saving for green networking is mainly realized by preventing cells from emitting at full power when there is no UE to serve.

- Interference reduction: Interference reduction can be performed through transmit power control and/or resource allocation. For instance, a LPN device scans the whole spectrum and chooses empty bands to transmit. Operators can also deploy their cells in orthogonal bands, reducing inter-cell interference with a low complexity but at the cost of reduced maximum performance.

- Automated configuration of PCI: In LTE, it is suggested that each cell should be able to automatically choose its cell identity. It is a challenging task due to the limited number of cell identities available, which may lead to identity collision or confusion. Therefore, gathering data from neighbouring cells to make a decision is important, especially in scenarios with large cell densities.

- Mobility robustness optimization: In LTE, small cells are required to optimize their handover parameters, assisted by previous handover-failure and ping-pong data.

- Mobility load balancing optimization: Small cells must be able to optimize cell re-selection and handover parameters to balance traffic load and minimize the associated number of handovers and redirections.

- Automatic Neighbouring Relation (ANR) function: A small cell must find ways to maintain an updated list of neighbouring cells.

- ICIC: Neighbouring cells collaborate through ICIC to allocate their transmit power and Resource Blocks (RBs) in a way that inter-cell interference is minimized.

- Random Access Channel (RACH) optimization: RACH settings need to be optimized to minimize interference among RACHs, and thus reducing the RACH collision probability, which affects call setup delays.

In 3GPP standards, parameters optimization is left up to the manufacturer/operator. Recent researches on SON techniques focus mainly on three approaches: cognition, game theory and learning, which are introduced in Section IX. Distributed SON techniques are more appealing to operators because they reduce deployment and maintenance costs, and may lead to near optimal solutions. Accordingly, a main challenge is to develop simple SON techniques that can be implemented on cheap devices with limited computing capabilities to provide near optimum network performance.

## IV. CARRIER AGGREGATION

LTE-Advanced, as an evolved version of LTE, set up more challenging performance targets [10]. One target of LTE-Advanced is to support transmission bandwidths of up to 100 MHz. This is achieved through CA, where multiple blocks of legacy LTE spectrum, i.e. CCs, are amalgamated to obtain a wider bandwidth [11] [12], while backward compatibility is maintained so that LTE-capable-only UEs can operate in an LTE-Advanced network and vice-versa. This is because each CC for LTE-Advanced CA is a legacy LTE carrier.

With the additional degree of freedom that arises with CA in the frequency domain, interference mitigation can be achieved by optimizing the allocation of available CCs among contending macrocell BSs and LPNs. CC selection methods, depending on the interference environment of BSs in an LTE-Advanced system, have been proposed in [13]-[16]. However, the methods in [13]-[15] require excessive signalling overhead among BSs. In [16], the proposed method reduces signalling overhead, but fails to offer explicit protection to cell-edge UEs in dense and uncoordinated deployments.

### A. CA Analysis and Techniques

In this section, a novel Dynamic Autonomous Component Carrier Assignment (DACCA) scheme is proposed, where each BS in the network adapts its CC usage to protect cell-edge UEs from detrimental DL inter-cell interference without compromising spectral efficiency. In order to enable DACCA, two types of CCs are defined according to their usage foreseen by BSs: Primary CCs (PCCs) and Secondary CCs (SCCs).

PCCs are used to protect cell-edge UEs from inter-cell interference so as to boost cell-edge capacity. The PCCs of a particular BS cannot be used by its interfering cells, which are identified using a pre-defined global Signal-to-Interference-plus-Noise Ratio (SINR) threshold $\gamma_{th}$, which represents the minimum tolerable SINR of UEs. In order to ban/block CCs at interfering cells, the BS send a PCC indicator to it interfering cells, which will then abstain from using the indicated CC. Thus, cell-edge UEs allocated in PCCs will experience low inter-cell interference.

SCCs are used to enhance spatial spectrum reuse. SCCs are orthogonal to PCCs, and cannot be blocked at any cells. SCCs can be allocated by cells to cell-centre UEs that face less interference, as long as they do not cause high interference to neighbouring cells.

The CC configuration in a cell remains unchanged within each time slot, but may be updated at the start of the next time slot. Each BS computes the CC assignment for the next time slot based on the feedbacks received from its connected UEs and neighbouring BSs regarding the previous time slot or even earlier ones.

The proposed DACCA has a low signalling overhead, since existing LTE signalling procedures are used, and the system reaches a stable point after only a few iterations.

Fig. 1 shows an example of DACCA, where BS C may cause interference to UE 1 served by BS A in $CC_1$ (as a PCC) and UE 3 served by BS B in $CC_2$ (as a PCC). Both $CC_1$ and $CC_2$ are thus blocked at BS C to avoid inter-cell interference. Similarly, BS A cannot use $CC_2$, while BS B cannot use $CC_1$. Since UE 2, served by BS A, is unlikely to be interfered by BS B or BS C, it can be allocated in $CC_3$ without causing high interference to UE 4 served by BS C in $CC_3$.

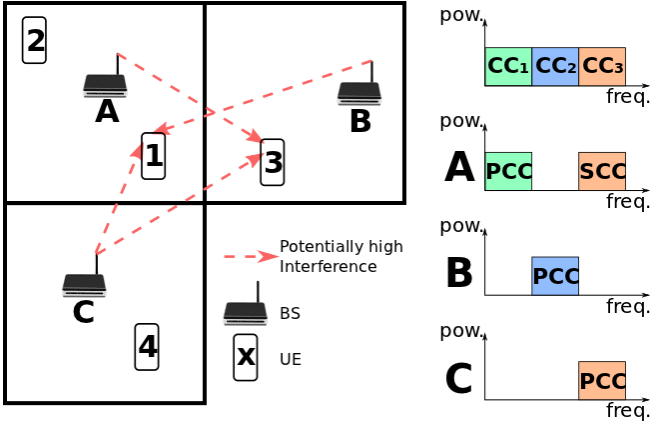

**Fig. 1** DACCA with 3 CCs contained in the system bandwidth.

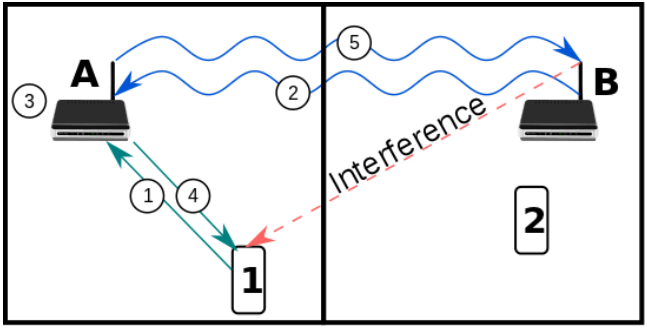

**Fig. 2** DACCA operation with 3 CCs in the system bandwidth.

Fig. 2 illustrates the operation of DACCA at BS A, where in step ①, BS A collects channel quality measurements from its UEs; in step ②, BS A receives PCC indicators from its neighbouring cells; in step ③, BS A updates its CC assignment for the next time slot based on the received feedbacks; in step ④, BS A allocates DL resources to its UEs according to the CC assignment; and in step ⑤, BS A sends a PCC indicator to its interfering cells. All BSs in the network perform the same DACCA operation independently. More detail on the allocation of CCs and RBs can be found in [17].

## B. Performance Evaluation

The performance of DACCA is benchmarked using the DL of an LTE-Advanced network with a total system bandwidth of 40 MHz, which is divided into 4 CCs of 10 MHz each. Each CC is comprised of a fixed number of RBs, which are the minimum DL resource allocation units in LTE-Advanced. A BS can assign RBs of the same CC to multiple UEs, but an RB can be assigned to at most one UE in a cell.

System-level simulations are performed using a one-story building scenario, modelled by a 5×5 grid, according to 3GPP specifications [34]. For the sake of simplicity, inter-cell interference from macrocells to femtocells is neglected, which may be accomplished by allocating orthogonal frequency bands to macrocells and femtocells. The presence of a CSG femtocell in an apartment block is governed by the probability of CSG femtocell deployment, and CSG femtocells are uniformly distributed inside apartments. UEs are uniformly distributed inside apartment blocks containing a femtocell and are forced to connect to it. A full buffer traffic model for UEs is considered, and throughput calculations are derived from the effective SINR of each scheduled UE by using a truncated Shannon bound approach. More details on the simulation scenario and parameter setting can be found in [17].

The performance of DACCA is compared to that of a Fixed Frequency Reuse (FFR) scheme where one BS is assigned one or two out of four available CCs, i.e. FFR 1/4 and FFR 2/4, respectively. For DACCA, in the first time slot, BSs randomly assign one CC as PCC and subsequently update their PCC and SCC assignment according to the proposed scheme. The SINR threshold $\gamma_{th}$ is set to 5 dB.

Fig. 3 plots the Cumulative Distribution Functions (CDFs) of the achieved DL SINR for the three considered schemes. With DACCA, nearly all UEs achieve an SINR exceeding $\gamma_{th}$=5 dB, thus providing fairness. However, the network achieves the best average SINR performance when femtocells use FFR 1/4, with which inter-cell interference is mitigated at the cost of reduced spectrum efficiency.

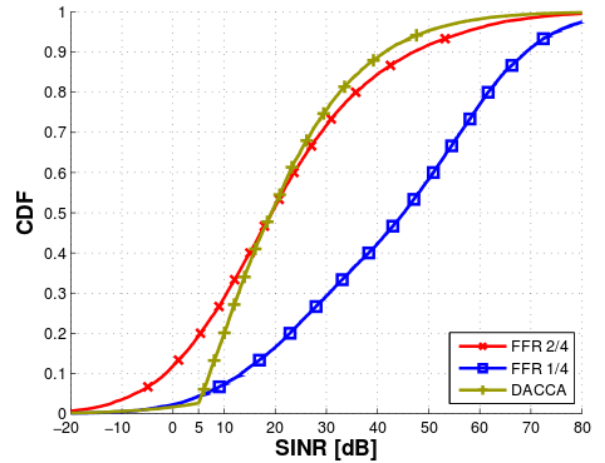

**Fig. 3** CDF of UE SINR.

**Fig. 4** shows the CDFs of the achieved UE capacity for the three schemes. Despite the encouraging SINR performance of FFR 1/4, it presents the worst UE average capacity due to the reduced spectrum efficiency. FFR 1/2 doubles the bandwidth for each cell and thus doubles the UE average capacity with respect to FFR 1/4, but worsens cell-edge performance. This is a typical trade-off in FFR: the less bandwidth per cell, the better SINR and cell-edge performance, but the worse UE average capacity. On the contrary, DACCA outperforms both FFR schemes, presenting a promising cell-centre performance in terms of very high average UE capacity, while showing an adequate cell-edge performance.

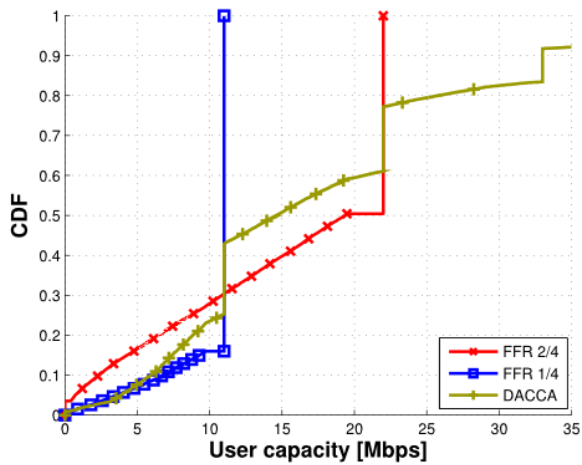

**Fig. 4** CDF of UE Capacity.

## V. ENHANCED INTER-CELL INTERFERENCE COORDINATION

New cell selection procedures that allow a UE to connect to the BS with the lowest path loss regardless of its DL Received Signal Strength (RSS) are being developed for a better spatial reuse. An approach under investigation is range expansion [5], in which a positive offset is added to the DL RSS of picocell pilot signals to increase their DL coverage footprint. Although range expansion mitigates UL cross-tier interference, it comes at the expense of reduced DL signal quality for Expanded Region (ER) picocell UEs (PUEs). The ER PUEs may suffer from DL SINRs below 0 dB, since they may connect to cells that do not provide the strongest DL RSS. Therefore, novel eICIC schemes for macrocells to cooperate with ER picocells, are required to mitigate excessive DL inter-cell interference for range expansion [18]-[20].

Within 3GPP, eICIC approaches based on ABSs have been proposed to mitigate cross-tier interference in range expansion scenarios. ABSs are LTE subframes that contain reference signals only, instead of control or data signals. As explained above, without inter-cell coordination, ER PUEs may observe high DL cross-tier interference from a macrocell, which can be mitigated by using ABSs at the macrocell and scheduling ER PUEs within the subframes that overlap with the ABSs of the macrocell. However, using ABSs may degrade the overall macrocell performance, since a given number of subframes are kept unused in order to achieve interference mitigation.

### A. Proposed eICIC Technique

In this subsection, a macro-pico coordinated RB and Power Allocation (coRPA) scheme that deals with DL macro-to-pico interference and enhances the macrocell performance, is proposed. The idea is that when a UE enters or stays within the ER of a picocell, the pico BS will inform the macro BS of the set of RBs allocated to this ER PUE, and then the macro BS will lower its transmit power in the specified RBs so that a desired DL SINR is guaranteed to this ER PUE. Inter-cell communication between macro BSs and pico BSs could be periodic or event triggered through the operator's backhaul network. The joint RB and power assignment scheme among macro BSs and pico BSs will tend to allocate macro UEs (MUEs) that are closer to the macro BS or have lower data-rate demands (therefore requiring lower transmit powers) to RBs that are used by ER PUEs. In this way, MUEs and ER PUEs can reuse the same RBs, while satisfying their respective SINR requirements, thus improving the macrocell performance as compared to the ABS approach.

The coRPA scheme can be implemented in two steps:

- First, decide the maximum power that the macro BS can apply in each RB being used by ER PUEs, in order to guarantee the desired DL SINR to ER PUEs. This requires some form of coordination/communication via message passing between macro BSs and pico BSs.

- Second, the macro BS allocates RBs and transmit powers to its DL UEs, while respecting the maximum power constraints derived in the first step.

For more detail about how transit power constraints are computed by the macrocell and picocells, and how RBs are allocated to UEs following the proposed coRPA scheme, the reader is referred to [21].

### B. Performance Evaluation

In this subsection, performance of the proposed macro-pico coRPA scheme in support of ER picocells is evaluated through system-level simulations in a scenario that comprised 1 macro BS and 2 or 4 outdoor pico BSs. The outdoor pico BSs and MUEs were uniformly distributed around the macro BS within its coverage radius, while PUEs were uniformly distributed within a 40 m hotspot radius around each pico BS. The minimum distance between a macro BS and a pico BS was 75 m, and the minimum inter-pico-BS distance was 40 m. Path loss models and other parameters were selected according to 3GPP recommendations for outdoor picocells (model 1) in [14]. More information of the system-level simulation is provided in [21].

Network performance is assessed in terms of three Key Performance Indicators (KPIs):

- UE outages: the total number of outages incurred to UEs during the entire simulation;

- Connected UEs: the average number of UEs connected to the network simultaneously along the entire simulation;

- Network throughput: the average sum throughput carried by the network during the entire simulation.

Moreover, the following schemes are included in the system-level simulations for performance comparison:

- Uniform Power Distribution (UPD): Each macro/pico BS uniformly distributes its DL transmit power among all subcarriers, targeting a frequency reuse factor of 1, and allocates RBs in each cell independently.

- UPD + Resource Partitioning (RP): Each macro/pico BS uniformly distributes its DL transmit power among all subcarriers. Half of the RBs are used by the macrocell, and the other half are used by the picocells, thus avoiding cross-tier interference between macro- and pico-cells.

- coRPA with macrocell-picocell cooperation: Each pico BS uniformly distributes its power among subcarriers, and the macro BS employs the coRPA scheme, with power constraints in the macrocell resource allocation.

All the above schemes were tested for different ER offsets, i.e. $\Delta ER_p = \{0, 8, 16\}$ dB, where $\Delta ER_p = 0$ dB means no ER.

The results of our system-level simulations in terms of the three KPIs are presented in Tables I and II for the scenarios with 2 and 4 picocells, respectively.

**Table I** Comparison of average performance with 2 picocells.

| Technique | KPI | UPD | UPD + RP | coRPAP |
|---|---|---|---|---|
| $\Delta ER_p = 0$ dB | outage | 93 | 135 | 87 |
| | users | 49.89 | 28.14 | 53.24 |
| | Mbps | 14.685 | 8.394 | 15.849 |
| $\Delta ER_p = 8$ dB | outage | 99 | 96 | 49 |
| | users | 46.69 | 50.41 | 75.64 |
| | Mbps | 14.814 | 16.010 | 23.782 |
| $\Delta ER_p = 16$ dB | outage | 103 | 52 | 0 |
| | users | 44.98 | 74.26 | 100 |
| | Mbps | 12.961 | 24.171 | 31.986 |

**Table II** Comparison of average performance with 4 picocells.

| Technique | KPI | UPD | UPD + RP | coRPAP |
|---|---|---|---|---|
| $\Delta ER_p = 0$ dB | outage | 188 | 219 | 172 |
| | users | 50.07 | 34.48 | 59.52 |
| | Mbps | 14.609 | 10.777 | 17.920 |
| $\Delta ER_p = 8$ dB | outage | 187 | 132 | 81 |
| | users | 49.60 | 85.15 | 110.72 |
| | Mbps | 14.710 | 28.370 | 35.698 |
| $\Delta ER_p = 16$ dB | outage | 217 | 71 | 0 |
| | users | 40.85 | 118.87 | 150 |
| | Mbps | 12.765 | 39.642 | 49.002 |

In both tables, we can see that the average number of network connected UEs increases with the ER offset. Without ER (i.e. $\Delta ER_p = 0$ dB) or for a small ER (e.g., $\Delta ER_p = 8$ dB), a picocell cannot serve all hotspot UEs around the pico BS, and hence the macrocell is overloaded with more UEs than it can support. Such macrocell overload translates into a large number of UE outages. On the contrary, for $\Delta ER_p = 16$ dB, all hotspot UEs around a pico BS can connect to it, thus providing load balancing and a much better spatial reuse.

For $\Delta ER_p = 16$ dB in Table II, we also make the following

observations:

- When using UPD, a large number of outages occur and only 40.85 out of 150 UEs on average can connect to the network due to the large inter-cell interference suffered by ER PUEs from the umbrella macrocell.

- When using UPD+RP, the number of average network connected UEs increases to 118.87, because RP fully removes cross-tier interference. However, outages still occur, because only 25 out of the 50 mobile UEs can connect to the macrocell, and mobile UEs may not be able to handover from the macrocell to the picocells, as a result of RP.

- When using the proposed coRPA scheme, there was no UE outage observed at all, and an average of 109.15 more UEs were simultaneously connected to the network than using the non-cooperative UPD approach. This is because based on the information exchanged between the picocells and the macrocell, the macro BS was able to allocate mobile UEs requesting low DL transmit power in the RBs being used by ER PUEs, allowing for a better spatial reuse that increased the average network throughput by more than 3 times with respect to UPD and 23.61% with respect to UPD+RP.

## VI. MULTIPLE INPUT MULTIPLE OUTPUT

In LTE-Advanced, CA increases system capacity by using more spectrum, and eICIC mitigates inter-cell interference by intelligent resource allocation. Moreover, spatial multiplexing offered by MIMO techniques is also explored.

Performance enhancement achieved by MIMO in LTE homogeneous macrocell networks has been demonstrated in [23]-[26]. In [27], the performance of single-user MIMO (SU-MIMO) in LTE was analyzed, assuming frequency reuse 1 in both macrocell and femtocell networks with a maximum of two femtocells per macrocell. In [28], analytical models were developed to evaluate the coverage of a macro-femto two-tier network employing SU-MIMO and multi-user MIMO (MU-MIMO), assuming flat Rayleigh fading only per sub-band for analytical simplicity. While in [22], it was shown that MIMO performance would be affected by inter-cell interference.

### A. Analysis and Proposed Techniques

In this subsection, we evaluate the performance of 3GPP LTE Rel-8 macrocell-femtocell deployments when SU-MIMO and FFR are used at macrocells through system-level Monte-Carlo simulations, which account for not only fast fading (using the MIMO Spatial Channel Model Extended (SCME) of [35]) but also path loss and lognormal shadowing (according to model 1 of [34]). FFR leads to the availability of orthogonal sub-bands in the coverage area of a macrocell, which can be exploited by the underlaid femtocells. Decisions on which sub-bands to use in each femtocell could be made either by a central entity or by the femtocell autonomously. Similar to [29], we consider a distributed approach, where sub-band assignments are self-organized at femtocells, while an FFR scheme is applied at macrocells.

### 1) Macrocells

In order to analyse FFR for LTE-Rel. 8/9 specifications, we use the macrocell layout comprising 7 sites with 3 sectors per site as illustrated in Fig. 5, in conjunction with wraparound. In Fig. 5, each cell is divided into an inner region and an outer region by a line that is perpendicular to the cell bore sight direction (at a distance $R_0$ from the BS). Sub-bands in the set $S_0$ are allocated to the inner regions of all cells for frequency reuse 1, while sub-bands in the three orthogonal sets $S_1$, $S_2$ and $S_3$ are allocated to the outer regions of all cells for frequency reuse 3, where $S = S_0 + S_1 + S_2 + S_3$ and $|S_1| = |S_2| = |S_3|$.

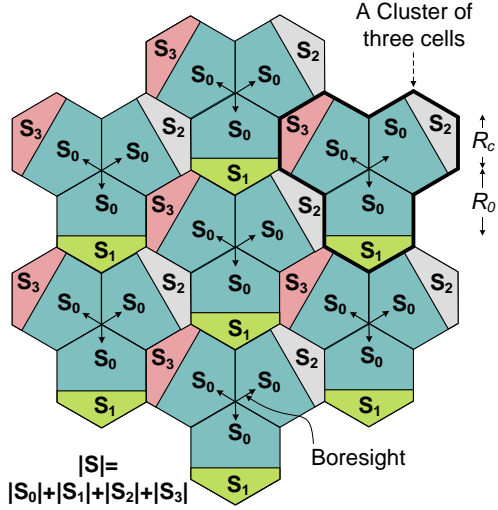

**Fig. 5** FFR scenario in the macrocell network.

### 2) Femtocells

In every macrocell, there will be $S - S_0 - S_m$ ($m \in \{1, 2, 3\}$) sub-bands left unused, which can be used by the underlaid femtocells. Moreover, femtocells located at cell-edges can also use the sub-bands in $S_0$ that are used only by the macrocell inner regions. Femtocells in the network monitoring mode (see Section III) are assumed to be able to measure the RSSs of all sub-bands in $S$ from neighbouring macro and femto-cells [33]. Accordingly, each femtocell transmits in the sub-bands in which it receives the least interference.

### B. Performance Evaluation

In the simulations, we consider four MIMO configurations: 1×1, 2×2, 4×2 and 4×4. Each UE has a velocity of 3 km/h. A Minimum Mean Square Error (MMSE) receiver is applied on each subcarrier. SINR per subcarrier is calculated over every spatial layer transmitted, and the effective SINR over multiple subcarriers is computed using the Mean Instantaneous Capacity (MIC) model [36]. Throughput is calculated from the effective SINR of each scheduled UE using a truncated Shannon bound.

Femtocells are deployed according to the dual-stripe model [34], where one dual-stripe cluster is randomly dropped in each macrocell. Each stripe has 6 floors, and each floor contains 40 apartment blocks of 10 m×10 m each. There is a probability of a CSG femtocell uniformly distributed in an apartment block. MUEs are uniformly distributed within the macrocell coverage area, such that a certain number of MUEs are attached the macrocell according to the best link criteria. Femto UEs (FUEs) are uniformly distributed inside apartment blocks containing a CSG femtocell and connect to it. A full buffer traffic model is considered for both MUEs and FUEs.

UEs report measured Channel Quality Indicators (CQIs) and/or Pre-coding Matrix Indicators (PMIs) every 5 Transmission Time Intervals (TTIs), and Rank Indicators (RIs) every 10 TTIs, for all sub-bands to their serving BS, indicating the combination of PMI and RI that delivers the maximum throughput. The reported RI is based on wideband measurements and indicates the number of layers a UE can support. Based on the received reports, a macro BS allocates its resources to the UEs considering also scheduling fairness. More detail on the allocation of CCs and RBs can be found in [37].

The average cell throughputs for the four considered MIMO configurations are given in Table . FFR used in macrocells decreases the average macrocell throughput by 12.5% and 9.5% for the 1×1 and 4×4 configurations, respectively, relative to frequency reuse 1. Hence, the loss of average macrocell throughput due to FFR reduces with the increase of antennas used for MIMO. Due to the reduction of cross-tier interference offered by macrocell FFR, FFR used in macrocells provides an average femtocell throughput gain, which is higher for a higher-order MIMO configuration. The average UE spectral efficiencies for the four considered MIMO configurations are also given in Table . FFR used in the macrocells increases the spectral efficiency of macrocells by 14.3% and 24% for the 1×1 and 4×4 configurations, respectively, as compared to frequency reuse 1. Similar results are also observed for the femtocells. Overall, interference mitigation yields a higher gain in terms of average UE spectral efficiency and average cell throughput for an LTE macro-femto network employing MIMO as compared to Single Input Simple Output (SISO). The network monitoring mode based sub-band selection in a femtocell is able to well protect MUEs trapped in the vicinity of a femtocell.

**Table III** Comparison of average values of KPIs.

|  | MIMO configuration | Macrocell values | | Femtocell values | |
| --- | --- | --- | --- | --- | --- |
|  |  | Macro reuse 1 | Macro FFR | Macro reuse 1 | Macro FFR |
| Average cell throughput (Mbps) | 1×1 | 16 | 14 | 6.7 | 8 |
|  | 2×2 | 25.1 | 22 | 11.6 | 14.1 |
|  | 4×2 | 29.8 | 26.1 | 13.2 | 15.7 |
|  | 4×4 | 43 | 38.9 | 21 | 26 |
| Average spectral efficiency (bps/Hz) | 1×1 | 1.4 | 1.6 | 3.3 | 3.7 |
|  | 2×2 | 2.4 | 2.9 | 5.8 | 6.6 |
|  | 4×2 | 2.8 | 3.4 | 6.6 | 7.3 |

| | 4×4 | 4.2 | 5.2 | 10.5 | 12.1 |
|---|---|---|---|---|---|

## VII. Green Small Cells: New Architectures

Energy efficiency is both ecologically and commercially important to Information and Communication Technologies (ICT). Over 0.5% of the global energy consumption comes from wireless communication systems, mainly by outdoor cellular network BSs [38]. A key challenge is to significantly reduce the energy consumption level whilst maintaining and even enhancing network capacity. Moreover, in order to improve competitiveness and the average revenue per UE, operators have to reduce operational cost of cellular networks.

Existing research on reducing the energy consumption of cellular networks has mainly focused on capacity improving transmission and RRM techniques, such as MU-MIMO and CoMP. Considering the total energy consumption of the Radio Access Network (RAN), the amount of energy saved by transmission and RRM techniques alone is fundamentally limited, while the energy saved by re-deployment can be much more significant [39].

The relationship between spectral and energy efficiencies has been characterized in [40] for noise-limited channels, but it is largely unexplored for interference-limited channels [41]. The cost efficiency of HetNets has been considered in [42], but the relationship between spectral, energy and cost efficiencies has not been established, especially in the context of interference-limited networks.

### A. Proposed Analysis and Techniques

In this subsection, the 3-way trade-off between spectral, energy and cost efficiencies for a variety of target throughputs is studied. Moreover, we investigate the energy efficiency of LPN networks relative to a reference macrocell network.

We consider an LTE network comprising multiple outdoor BSs and UEs, with two different deployment scenarios:

- Reference: a low density of high-power and high-cost 3-sector microcell BSs;
- Small cell: a high density of low-power and low-cost 1-sector picocell BSs.

For a UE served by BS $i$, the SINR per sub-carrier is

$$\text{SINR} = \frac{P_{\text{TX},i} H_i A_{\theta i} \lambda d_i^{-\alpha}}{\sum_{j \in N, j \neq i} P_{\text{TX},j} H_j A_{\theta j} \lambda d_j^{-\alpha} + n}, \quad (1)$$

where $P_{\text{TX},i}$ is the transmit power per subcarrier of BS $i$, $d_i$ is the distance to BS $i$, $\alpha$ is the path loss exponent, $\lambda$ is the path loss constant [43], $H_i$ is channel gain from BS $i$ taking into account effects of fading and lognormal shadowing, and $A_{\theta i}$ is the antenna gain [43].

#### 1) Spectral Efficiency

The spectral efficiency of a system is defined as the system-level throughput that can be achieved per unit of bandwidth (bit/s/Hz). The system-level throughput has been obtained from simulations using adaptive Modulation and Coding Schemes (MCSs) [43]. An analytical expression of spectral efficiency based on a modified Shannon expression that accounts for modulation, capacity saturation and Forward Error Correction (FEC) coding imperfections is given by

$$\eta_{\text{SE}} = \min\{\log_2\left(1 + \frac{\text{SINR}}{1.5}\right), 4.2\}, \quad (2)$$

where the factor 1.5 accounts for coding imperfections, and the capacity saturation of an LTE transmission is typically 4.2 bit/s/Hz.

It is also worth defining the load $L$ of a BS as a function of the offered traffic rate and the achievable throughput of the BS, i.e. $L = \frac{R_{\text{traffic}}}{B\eta_{\text{SE}}}$, where $R_{\text{traffic}}$ is the traffic rate offered to the BS, and $B$ is the available bandwidth.

#### 2) Energy Efficiency

The energy efficiency of a system is defined as the system-level throughput achieved per unit of power (bit/s/W or bit/J). The power consumption of a BS is typically a function of the load-dependent radio-head and load-independent over-head [44], and can be modelled as follows

$$P_{\text{BS}} = N_{\text{a}}(\frac{P_{\text{TX}}}{\mu}L + P_{\text{OH}}), \quad (3)$$

where $N_{\text{a}}$ is the number of antennas per BS, $\frac{P_{\text{TX}}}{\mu}L$ denotes the radio-head term, $\mu$ is the radio-head efficiency and $P_{\text{OH}}$ is the overhead term, which includes power consumption attributed to baseband processing, cooling, and backhaul [44]. The energy efficiency is then given by $\eta_{\text{EE}} = \frac{\eta_{\text{SE}}}{P_{\text{BS}}}$.

#### 3) Cost Efficiency

The Operation and Maintenance (O&M) cost efficiency of a system is defined as the system-level throughput achieved per unit of cost (bit/s/$). The annual cost expenditure of a BS is typically a function of the power consumption $P_{\text{BS}}$ and the rental costs, and can be modelled as follows

$$C_{\text{BS}} = P_{\text{BS}} T C_{\text{electricity}} + C_{\text{rental}}, \quad (4)$$

where $C_{\text{electricity}}$ is the cost of electricity in ($/kWh), $T$ is the number of hours that the BS is active over a year, and $C_{\text{rental}}$ is the rental cost associated with site and backhaul [41] [42]. The cost efficiency is then given by $\eta_{\text{CE}} = \frac{\eta_{\text{SE}}}{C_{\text{BS}}}$.

### B. Performance Evaluation

Fig. 6 compares different small cell deployment strategies in terms of spectral, energy, and cost efficiencies. We can see that the highest spectral efficiency is achieved with a high-density deployment of femtocells, which improves spatial reuse but creates a high level of inter-cell interference, thus diminishing the energy efficiency. The best energy efficiency is achieved with a medium-density deployment of picocells (each with a 200 m radius), while the highest cost efficiency is achieved with a lower but similar density deployment of picocells (each with a 230 m radius).

Fig. 6 also shows that, compared to the reference scenario, the small cell network that comprises a denser deployment of lower-power and lower-cost cells can save energy (by around 30%), but increases cost (by around 14%). This is primarily because the backhaul rental cost for a high density deployment is too high. For more detailed discussions about trade-offs between spectral, energy and cost efficiencies in deployments of LPNs, and the related theoretical analysis and simulations, the reader is referred to [41].

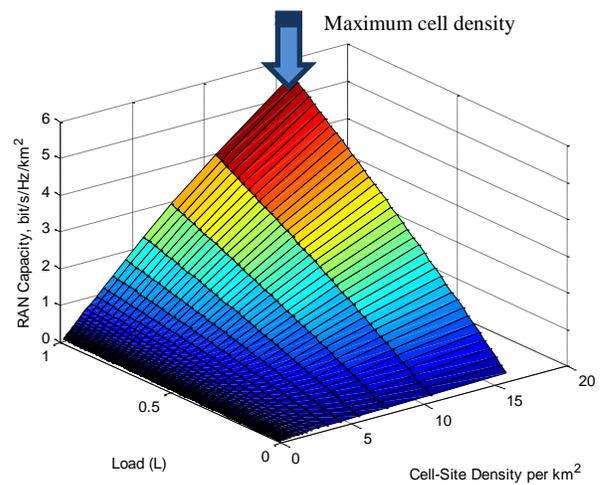

(a) Spectral efficiency vs. cell density and load.

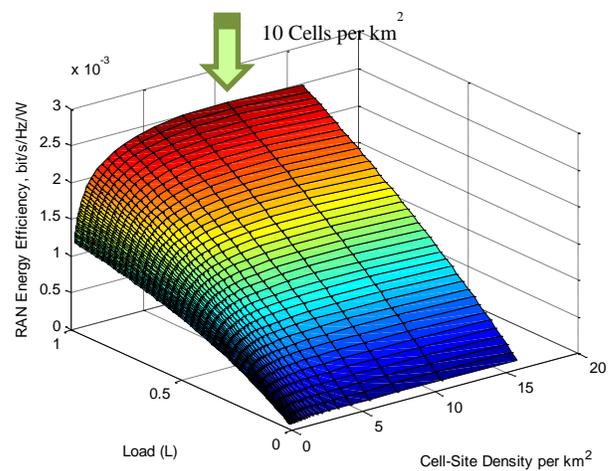

(b) Energy efficiency vs. cell density and load.

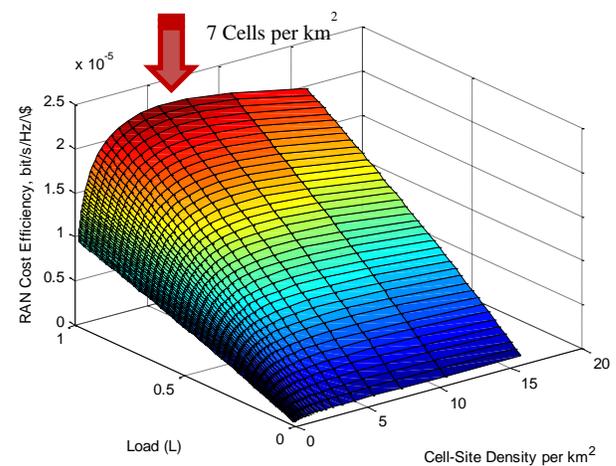

(c) Cost efficiency vs. cell density and load.

**Fig. 6** Small cell network **(a)** spectral efficiency, **(b)** energy efficiency, and **(c)** cost efficiency, as a function of deployment density and traffic load.

## VIII. GREEN RADIO RESOURCE MANAGEMENT

Forecast on the wireless market assumes continuous increases in subscribers, data rates per subscriber, and BSs for next generation mobile networks. The undesired consequence is the growth of wireless network energy consumption that increases operators' operational cost and adds to the global carbon dioxide emissions. Energy efficiency has become an alarming bottleneck in the growth paradigm of telecommunications.

According to a recent survey [45], nearly 80% of the energy consumption of a typical cellular network comes from the BSs. Furthermore, 70% of the BS energy consumption is caused by power amplifiers and air conditioning, which are used to keep the BS active even when there is no traffic in the cell. Hence, the optimization of radio access for BSs should have a large impact on the overall cellular energy efficiency.

Information theory based energy-efficient transmission schemes [46] [47], and solutions trading off spectral and energy efficiencies [48] or delay [49] for energy saving have been proposed. However, considering only Radio Frequency (RF) radiated power may lead to misleading conclusions. Novel RRM schemes should take into account the characteristics of RF front-end such as power amplifiers as well as UL and DL power and bandwidth constraints. Auer *et al.* [50] provided an estimation of the power consumption of several BSs considering different components of the radio equipment, such as antenna interface, power amplifier, baseband interface, cooling system, etc.

Although cellular traffic load notably varies during the day, mobile operators deploy BSs to accommodate the peak traffic demand. Classic RRM algorithms, which aim to maximize the system capacity while overcoming the mismatch between requested QoS and limited network resources under full system load, are not necessarily efficient under all kinds of operating conditions. Effective macrocell traffic offloading may reduce energy consumption at the macro BS. Originally envisioned as a means to provide better coverage and higher data rate in a given region, small cells are now primarily viewed as a cost-effective way to offload data traffic from the macrocell network.

The power consumption estimation in [50] can be used to evaluate energy-efficiency benefits of small cell deployments in cellular networks. Accordingly, the required input power $P_{in}$ to attain a certain RF output power $P_{out}$ can be computed as

$$P_{in} = \begin{cases} P_0 + \Delta_p P_{out}, & 0 < P_{out} \leq P_{max} \\ P_{sleep}, & P_{out} = 0 \end{cases} \quad (5)$$

where $P_{max}$, $P_0$, and $P_{sleep}$ are the RF output power levels at the maximum load, the minimum load, and in sleep mode, respectively, and $\Delta_p$ represents the dependency of the required input power on the traffic load.

The model in equation (5) indicates that

• Macro BS power consumption strongly depends on the traffic load, and thus macrocell traffic offloading to small cells can greatly enhance the overall cellular energy efficiency;

• Although LPN power consumption is less dependent on

traffic load, energy efficiency can be improved by adaptively switching the LPN off when it is not serving any active UE.

Note that massive and unplanned roll out of LPNs may also drastically increase the overall cellular energy consumption. The impact of small cell density on the network energy efficiency was studied in [51]. Ashraf et al. [52] investigated energy saving procedures that allow a LPN to dynamically deactivate/activate transmission functionalities according to the presence/absence of UEs in its coverage area. Frenger et al. proposed cell Discontinuous Transmission (DTX) [53] to enable BSs to switch off radio operations in subframes with no data transmission (see Fig. 7 (a)), saving energy mainly in low traffic scenarios. In an extended version of cell DTX, i.e. E-DTX [54], the UE data is buffered and transmitted as much as possible during the transmit intervals (see Fig. 7 (b)). Thus, the BS exploits the available frequency resources more efficiently and introduces longer silent intervals, at the cost of possibly higher delays in the application layer. Due to the limited number of UEs that can be simultaneously served by a LPN and the short distances between the LPN and its UEs, the spectrum resource is often under-utilized at small cells. Hence, E-DTX is a promising technique in small cell deployments.

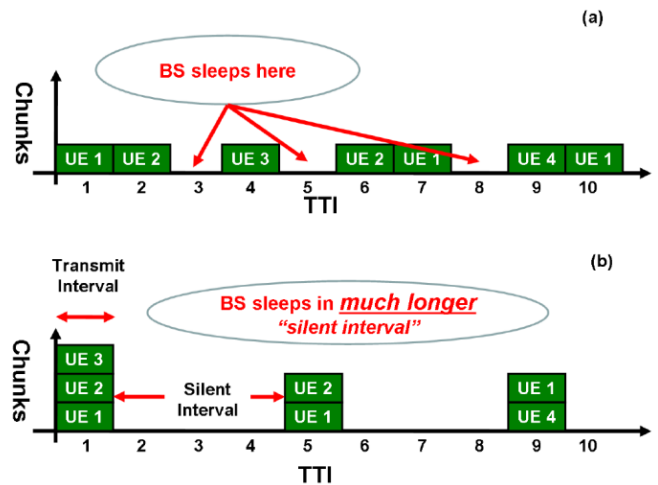

**Fig. 7** (a) Classic cell DTX, and (b) E-DTX schemes.

The more data transmitted in the activated TTIs, the higher the probability of introducing sleep intervals in future TTIs. Targeting moderate traffic load scenarios, a multi-cell DTX (MC-DTX) scheme [69] was proposed to adaptively distribute UE data among neighbouring small cells and control their activation/deactivation, in order to reduce the system power consumption and co-channel interference. Although a large number of LPNs is inter-connected via a high-speed low-latency backhaul, efforts are needed to keep packet delays below the QoS delay requirement. As shown in Fig. 8, the received traffic is classified into high and low priority ones according to the delay constraint. High priority traffic needs to be sent within the next few TTIs before the packets will be dropped by the user application, while low priority traffic has much less stringent constraint on delay.

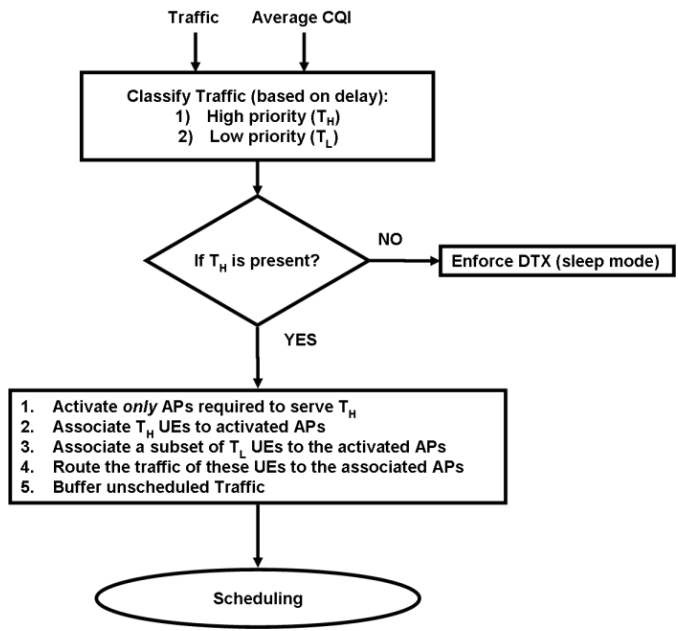

**Fig. 8** MC-DTX to manage traffic and LPN activity in cooperative small cell networks.

Although in both E-DTX and MC-DTX, the system buffers UE data to allow silent intervals as long as possible at serving LPNs, MC-DTX exploits inter-cell cooperation to adaptively associate high-priority and low-priority UEs to activated LPNs to increase the traffic sent in their transmit intervals. The performance of MC-DTX is compared with E-DTX through simulations in a femtocell network comprising both closed access and open access femtocells. In closed access femtocell deployment, a UE can be served only by the femtocell placed in the same apartment, the availability of which is controlled by a femtocell deployment ratio $\rho_d$. In open access femtocell deployment, UEs can be in the coverage areas of several femtocells, and each UE selects the femtocell providing the best link. The traffic of a femtocell UE is modelled as a Near Real Time Video (NRTV) traffic [70]. A proportional fair (PF) scheduler [71] is used at each femtocell.

Fig. 9 shows the average power consumption of E-DTX with closed access, E-DTX with open access, and MC-DTX with open access in the femtocell network versus the femtocell deployment ratio $\rho_d$ [72]. We can see that at low femtocell densities, the closed access deployment achieves lower power consumption of the femtocell network than the open access deployment. This is because the probability of a UE being served by a femtocell in the closed access deployment is lower (and thus longer achievable silent intervals) than that in the open access deployment. The power consumption offered by E-DTX increases with $\rho_d$, while the power consumption of MC-DTX nearly stops increasing when $\rho_d$ goes beyond 0.5. This indicates that MC-DTX is able to avoid energy wasting at medium to high femtocell densities by adapting femtocells' activities to the network topology and load through a dynamic UE-cell association. Overall, MC-DTX outperforms E-DTX by up to 50% in terms of power saving.

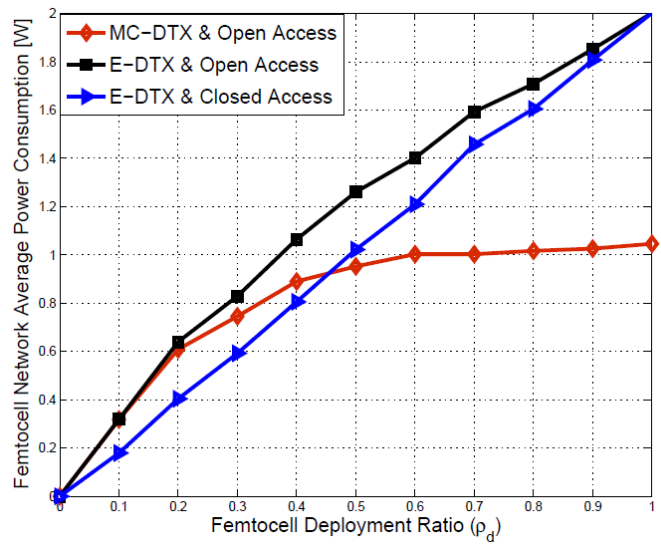

**Fig. 9** Femtocell network average power consumption versus $\rho_d$.

## IX. ENABLING DECENTRALIZED OPERATION

Centralized decisions are likely to be impracticable in heterogeneous small cell networks. LPNs should be able to self-organize into coherent behaviors in accordance with the environment. Self-organization is the ability of network nodes to dynamically adapt their operating parameters to improve individual or global performance. Self-organization thus requires the network nodes to collect information of the environment and perform the adaptation accordingly [55].

### A. Learning in Decentralized Systems

In self-organization, learning algorithms can be used to translate the sensed environmental information into actions. Machine Learning (ML) is a branch of Artificial Intelligence (AI) that involves the design and development of behavioral rules based on empirical information, such as data collected from sensors or past experience saved in databases [56].

For learning, decentralized systems are interpreted as multi-agent systems, which are especially useful in solving complex, large and unpredictable problems. Each agent in a multiagent system is specialized at solving a specific problem. A multiagent system is deliberative, if a model can be formalized for each agent behavior in terms of beliefs, desires and goals. A multiagent system is reactive, if the agents do not have an environmental representation and act in a stimulus-response manner [57]. Wireless systems, where problems are dynamic and interdependent (i.e. the system structure dynamically changes), are more adequately modeled as reactive agents. As a form to implement reactive agents, RL techniques can be applied. RL-based agents learn from observations of the environment when a given action is executed. There are three fundamental classes of methods to solve RL problems: dynamic programming, Monte Carlo methods, and Time Difference (TD) learning. Dynamic programming methods require a complete and exact mathematical model of the environmental dynamics. Monte Carlo methods are simpler than dynamic programming, but are not suitable for step-by-step incremental computation. TD learning methods do not require modeling of environmental dynamics and are fully incremental, but they are more complex to analyze than Monte Carlo methods [58].

## B. On-line Learning: A TD Learning Method Enhanced through Docition and Fuzzy Q-learning

In a scenario where multiple small cells underlay and share the same spectrum with macrocells, centralized frequency planning is not feasible given the decentralized nature of the network, but the Q-learning method (as a form of TD learning) can be used to solve the problem. In Q-learning, each small cell acts as an intelligent and autonomous agent that learns a power allocation policy to control, in a decentralized way, the aggregate interference they generate at nearby MUEs [59].

In a decentralized multiagent system, the environment perceived by an agent is no longer stationary, since other agents are adapting too. The learning process can thus have a high complexity (in terms of required operations and memory) increasing with the observation space. Docitive radio [60] was introduced as a possible solution to speed up the learning process and to create rules for unseen situations via knowledge exchange among learners. While cognitive ("cognoscere" in Latin) radios emphasize on learning, docitive ("docere" in Latin) radios focus on teaching. It capitalizes on the fact that some nodes may have acquired a pertinent knowledge for solving a specific problem and are thus able to teach other nodes on how to solve the same or similar problem. To apply docition, intelligent agents have to measure their expertise, which is stored in the Q-table, establish a relation pattern with the other agents in the system to find the potential entities to cooperate with, and decide the degree of cooperation and the moment(s) to execute the cooperative process. For example, in startup docitive radios, docitive agents teach their policies (i.e. Q-table values) to newcomers upon joining the network. The relation pattern is established based on the interference that small cells cause to the macrocell system.

A critical issue of Q-learning using lookup tables is that, the environmental states and the available actions have to be represented by discrete values and therefore, the use of thresholds is mandatory. This entails an important intervention of learning algorithm designers selecting the thresholds for state representation and setting the values of available actions. The sizes of the state and action sets directly affect the feasibility of knowledge representation, system adaptability and hence its performance. When the number of state-action pairs is large and/or the input variables are continuous, the memory required to store the Q-table and the required learning time may become impracticable large.

To facilitate continuous state and action representation without the need of near infinite Q-tables, the Q-learning algorithm has been improved by the combination of Fuzzy Inference System (FIS) and RL, resulting in a Fuzzy Q-learning (FQL) algorithm [61]. FQL offers a more compact and effective expertness representation mechanism, avoids subjectivity in selecting discrete states and actions, and speeds up the learning process by incorporating offline expert knowledge in the inference rules [61].

## C. On-line Learning with Incomplete Information

In order to avoid a large amount of signaling messages exchanged in the network, a desirable learning algorithm should adapt the transmit/receive configuration based only on local information and observation of the environment. In [62], every observation of the environment, e.g., interference level caused by macrocells, is used to directly update a probability distribution that small cells use to select their future actions. In [63], observations of the environment are used to estimate the achievable performance of each possible transmit/receive configuration, and the estimations are then used to update the probability distribution that small cells use to select their future actions. The simple variation of the RL paradigm in [63] allows the network to converge to Nash equilibrium that each small cell achieves the optimal performance given the transmit/receive configurations adopted by all the other cells. Nash equilibrium operating points are especially important in decentralized small cell networks [64]. Coarse correlated equilibrium operating points have also been studied for small cell networks [65]. Convergence of the above algorithms remains the main constraint on practical applications [66].

## D. Self-Organized Resource Allocation in Small Cells

In order to mitigate inter-cell interference in heterogeneous networks comprising macrocells and outdoor/indoor small cells, the self-organized resource allocation scheme [67] has the following two-step hierarchical process (see Fig. 10):

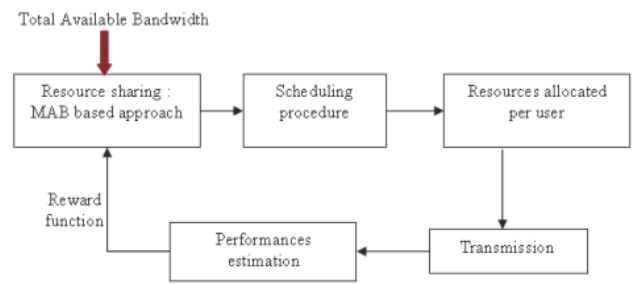

**Fig. 10** Proposed hierarchical scheme [67]

*Cell resources selection:* The total available frequency band is divided into a number of equal parts. Each cell autonomously selects the best band portion to transmit for a predefined period, based on an RL-based resource sharing scheme, which steers each cell to select the radio resources with the minimum interference level while ensuring reactivity to possible changes in resource usage. Each cell individually follows a Multi Armed Bandit (MAB) strategy to achieve the best performance in terms of UE interference level and throughput. The MAB strategy can be either deterministic or randomized. It provides a set of rules and policies for the cell to decide on the choices and actions for reaching a predefined objective.

*Per-user scheduling:* Each cell performs conventional scheduling to distribute its radio resources among attached

UEs. Each cell periodically evaluates the performance of the MAB resource allocation strategy, and deduces a reward function to be used for the next decision, based on the gain offered by a given MAB policy (e.g., the upper confidence bound algorithm [67] [68]).

The self-organized resource allocation scheme [67] is able to cope with variations in traffic and interference by adapting resource allocation accordingly in an autonomous manner. It is possible to integrate the self-organized resource allocation into existing radio access technologies with minor modifications, since it is independent of the scheduling implementation.

## X. SYSTEM-LEVEL SIMULATIONS

System-level simulations have been considered as an important tool to understand complex network behaviors and to predict system performance before the actual network deployment takes place, so as to avoid potential deployment pitfalls in the early design phase.

A comprehensive LTE system-level simulator where RRM and Medium Access Control (MAC) scheduling algorithms can be evaluated and tested has been developed in a joint project led by Ubiquisys and the Centre Tecnologic de Telecomunicacions de Catalunya (CTTC), based on the open source ns-3 network simulator [73]. Note that ns-3 is an Internet Protocol (IP)-based network simulator, which provides a lot of supports for higher applications. The ns-3 infrastructures can be inherited as the basic framework in the LTE simulator. The simulator is open to the community in order to foster early adoption, and provides a framework for Original Equipment Manufacturers (OEMs) and operators to test RRM strategies for macro, metro and femto deployment scenarios. Operators can use the simulator to test if large and small cells from different vendors can work together harmoniously before they are deployed.

The simulator aims to cover a relatively wide range of functionalities, but network security and RRM functionalities are left open for further developments and specific needs.

### A. Overall Architecture

The simulator contains two key components. The first one models the Enhanced Packet Core (EPC) of a 4G network, which consists of the core network, Serving Gateway (SGW), Packet Data Network Gateway (PDN GW or simply PGW), MME, and eNB. The second component models the radio interface of the network, which consists of the Radio Resource Control (RRC), Packet Data Convergence Protocol (PDCP), Radio Link Control (RLC), MAC, and the Physical (PHY) layer of the protocol stack. These protocol layers reside in the UE and eNB (see Fig. 11).

### B. EPC Model

Unlike many existing system-level simulators, which tend to focus only on the radio access aspect of the network, this simulator places also a strong emphasis on the EPC aspect of the network, thereby enabling the realistic modeling of end-to-end applications that rely on the interconnection of multiple

UEs to the Internet. Moreover, the simulator supports multiple bearers per UE. Essentially, the EPC model consists of two layers of IP networking: the first layer involves UE, PGW and remote host residing somewhere in the Internet; the second layer involves only eNB, SGW and PGW, where SGW and PGW reside in the same node for simplicity. IP packets associated with the first layer are tunneled through the second IP layer via the General Packet Radio Service (GPRS) Tunneling Protocol (GTP)/User Datagram Protocol (UDP)/IP in the S1-U interface (see Fig. 12).

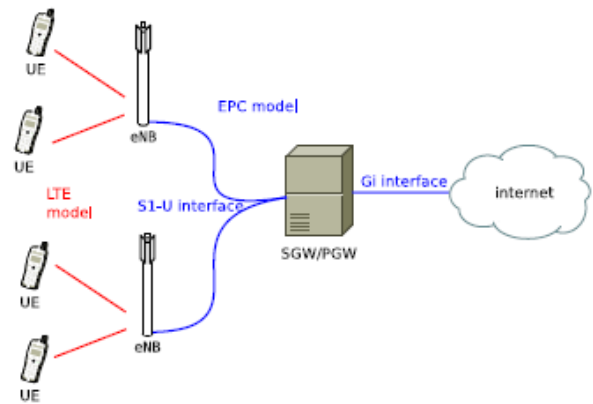

**Fig. 11** Overall architecture

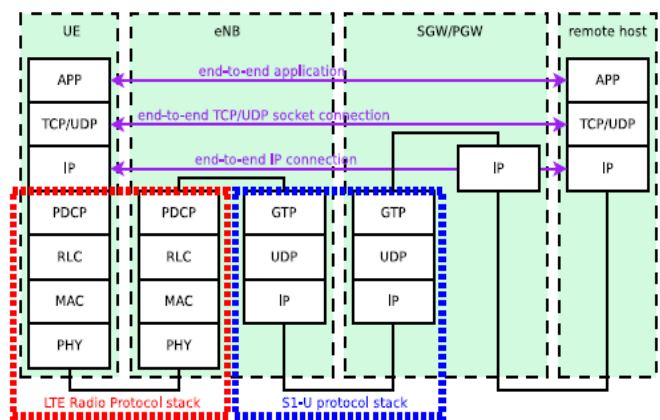

**Fig. 12** User plane protocol stack

### C. LTE Model

While the first layer IP networking is tunneled between the SGW/PGW and the eNB via the GTP/UDP/IP, it is the PDCP/RLC/MAC/PHY that is responsible for the IP packets between the UE and the eNB over the air interface. While it is possible to model the PHY layer at the symbol level, it is typically not practical to do so due to the associated high complexity. Instead, the link-level error model is abstracted in a look-up table, and an RB (rather than a symbol) is the smallest unit of simulation. Fast fading traces are first computed off-line, and then are read into the memory at run-time. Fig. 13 shows an example of the frequency selective fast fading in the Extended Pedestrian A model [76] for 3 km/hr.

The link-to-system mapping for the data plane is based on the Mutual Information Based Effective SINR Mapping (MIESM) [77]. In MIESM, the SINR of each RB is first calculated, the Mutual Information per Bit (MIB) per RB for each modulation is obtained, and the MIBs over all relevant RBs are averaged to obtain the Mean Mutual Information per Bit (MMIB). Finally, the transport block error rate is obtained based on the MMIB, the given transport block size (TBS), and the selected MCS (See Fig). Depending on the selected TBS, segmentation may be needed to break the transport block into a number of code blocks, and the corresponding code block error rate is obtained based on pre-computed link-level curves. After code block concatenation, the transport block error rate is calculated based on the code block error rate.

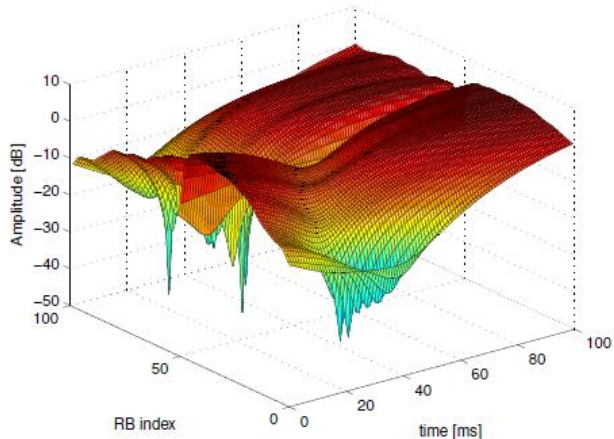

**Fig. 13** Frequency selective channel

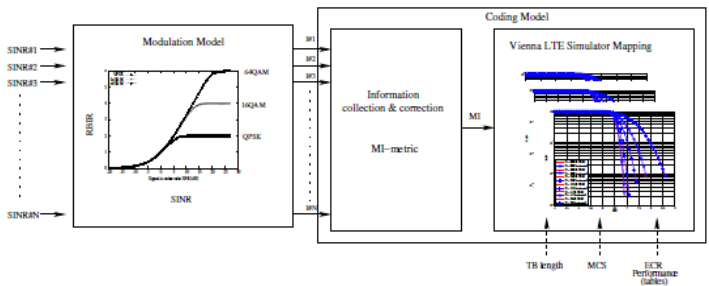

**Fig. 14** MMIB error model

The simulator takes into account both outdoor and indoor propagation models, including the Okumura Hata model, the short range ITU-R P.1411 model, indoor communications model I1238, etc. The appropriate model is invoked depending on the locations of transmitter and receiver. In addition, omni-directional and sectorized antennas can be incorporated in the same scenario simultaneously. This is particularly useful in scenarios involving both small cells and macrocells.

The RLC layer consists of the Transparent Mode (TM), Unacknowledged Mode (UM), and Acknowledge Mode (AM) [74]. In the current simulator, only UM and AM are supported, but a Simplified Mode (SM), which does not require the EPC and IP networking support, is introduced to model full buffer traffic in radio access only scenarios. Moreover, the current simulator supports only a simplified version of the PDCP layer [78], which is just enough to support data transfer. Due to the above simplifications, features such as header compression, in-sequence delivery of higher layer Protocol Data Units (PDUs), de/ciphering of user/control data are not supported.

# XI. SMALL CELL DEPLOYMENT SCENARIOS

## A. Static Installations

Femtocells are low-power BSs typically deployed indoors, e.g., in residential premises, enterprise office buildings, and hotspots. Due to the different characteristics of residential and enterprise femtocells, special care needs to be taken in their respective deployment, especially in co-channel deployments where femto and macro cells share the same spectrum.

Technical analysis and design optimization are essential in order to find the appropriate and optimized solution. It is also necessary for the design and optimization tool to be able to superpose and/or integrate with other techniques. Propagation prediction and system-level simulations must be performed to optimize the final solution. iBuildNet®, which is an in-building network planning & optimization software tool produced by Ranplan Wireless Network Design Ltd., offers a 3D modeling platform for design and optimization of femto/small cell installation.

Characteristics of residential femtocells include:

- Only a few users camped on each femtocell;
- Use of standard IP broadband as backhaul;
- Co-channel interference caused by femtocells to outdoor macrocells due to unplanned installation by subscribers.

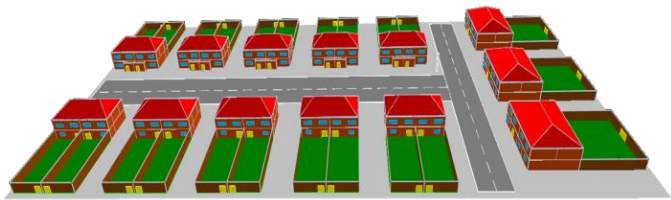

**Fig. 15** UK residential area scenario (screenshots generated in *iBuildNet®*).

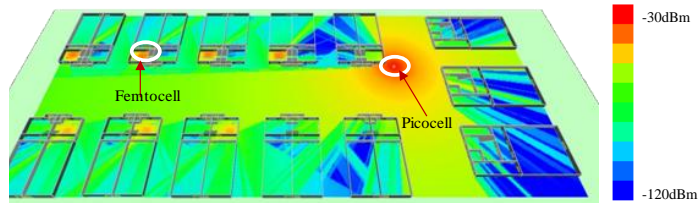

**Fig. 16** Best signal level for indoor and outdoor coverage.

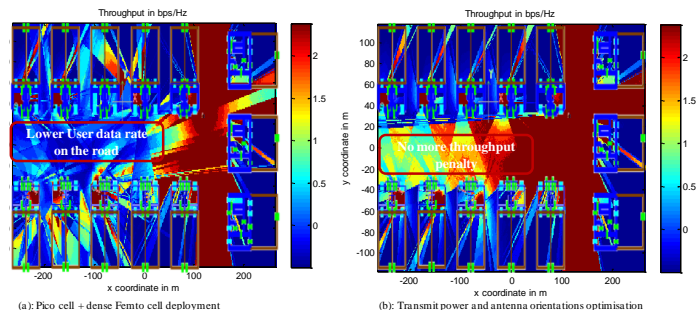

**Fig. 17** UE throughput w/o optimization.

A typical UK residential area is plotted in Fig. 15, where femtocells are deployed to ensure good indoor coverage within premises and a picocell is deployed to provide outdoor coverage. Fig. 16 depicts the best signal level for indoor and outdoor coverage. We can see that the signal is strong enough to meet the coverage requirements. However, indoor signal leakage causes strong interference to outdoor UEs, as shown in Fig. 17(a). Hence, residential femtocell deployment should focus on the mitigation of interference from femtocells to outdoor cells. Due to the unplanned nature of femtocell installation, low transmit power and antenna orientation at the femtocells are desirable. Fig. 17(b) shows that throughput per UE on the road has significantly improved when SON based on femtocell power control and antenna orientation adjustment is used to mitigate the interference.

Enterprise femtocells have the following characteristics:

- Inter-femtocell interaction;
- Larger number of users and coverage area;
- Higher uplink interference to surrounding macrocells due to the larger femtocell coverage;
- Femtocell-to-macrocell interference;
- Potentially large RF variations inside the building.

Fig. 18 depicts an office building where femtocells will be densely deployed. Fig. 19 plots the best signal level for indoor coverage on the 2nd floor. It can be seen that it is desirable to confine femtocell coverage within the building and lower the signal leakage outside. In Fig. 20(a), inter-femtocell interference is so severe that cell-edge UEs can obtain only low data rates. Fig. 20(b) shows that using multiple optimization techniques, such as femtocell transmit power optimization, femtocell location adjustment, and optimising the number of active femtocells, can mitigate the inter-femtocell interference.

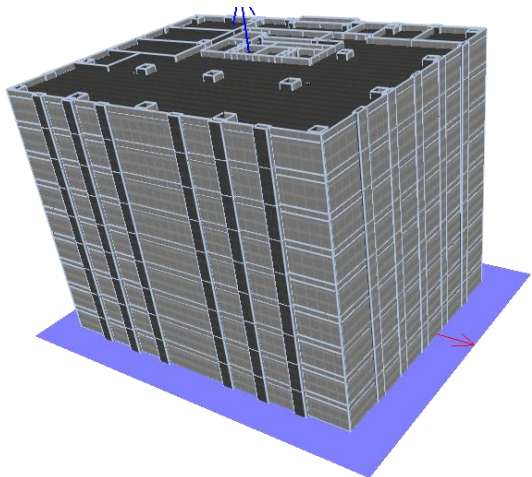

**Fig. 18** Office building scenario.

Fig. 21 presents the spectrum efficiency improvement of different optimization methods. Femtocell transmit power optimization can improve the spectrum efficiency by up to 17%. With a combination of these optimization techniques, the spectrum efficiency improvement is up to 39% as compared with that without optimization. It has been shown that unlike residential femtocells, enterprise femtocell deployments need network planning via proper femtocell placement and proper transmit power calibration towards the optimized coverage of all femtocells deployed. Accordingly, a larger number of femtocells with a smaller coverage each are desirable. The appropriate number of femtocells for a specific office building depends on the shape, size and materials of the building.

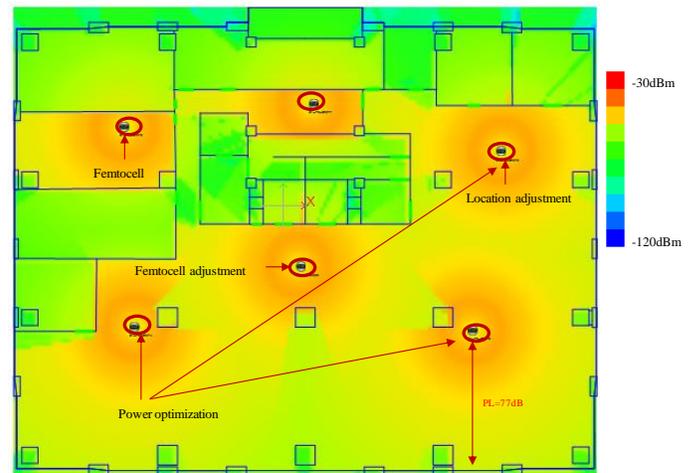

**Fig. 19** Best signal level for indoor coverage.

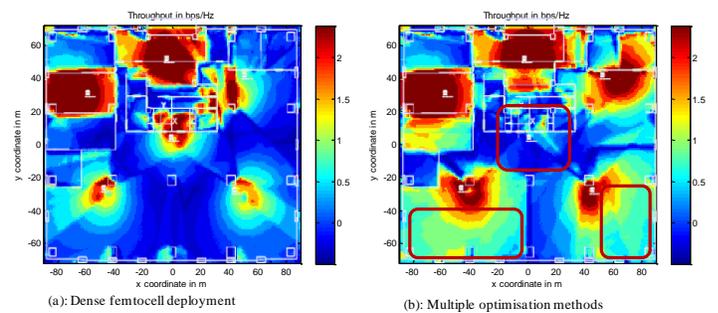

**Fig. 20** User throughput map between w/o optimization.

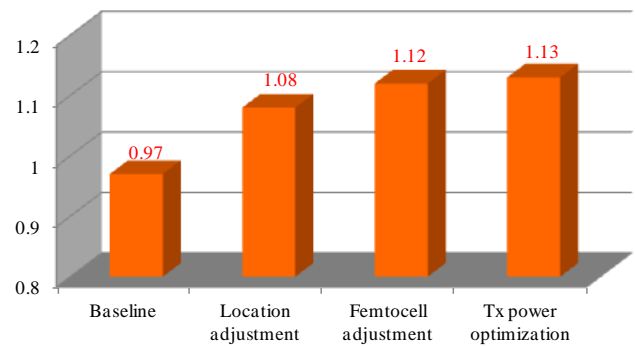

**Fig. 21** Spectrum efficiency comparison of different optimization methods.

## B. Mobile Installations

The deployment of small cells is not just limited to static locations on the ground. There are also nomadic and itinerary small cells. For example, picocells stored in depots and/or rapidly deployed in emergency situations are nomadic small cells [80]. First respondents may use nomadic small cells to quickly deploy a private mobile network in disaster affected areas. However, the radio interface of a nomadic cell is disabled during relocation. Picocells installed on board of cruise ships or aircrafts to provide coverage and services to onboard users are itinerary small cells, which have already found their way into commercial products especially in the aeronautical and maritime Global System for Mobile communications (GSM) markets [81]. The radio interface of an itinerary cell is active on the move.

Nomadic and itinerary small cells also require backhaul connections to the Public Switched Telephone Network (PSTN), Public Land Mobile Network (PLMN), or the Internet. Since these small cells are usually deployed in remote (e.g., desert, oil rig) or isolated (e.g., aircraft, ship) areas, their backhaul is currently mainly via satellite links that are the only communications technology available in remote or isolated regions [82]. Compared to conventional small cell backhaul methods such as broadband, the use of satellite airtime for backhaul is not cost effective at all. In addition, backhauling voice traffic over geostationary satellites is subject to a mouth-to-ear delay of approximately 500 ms, which is not acceptable according to the ITU G.114 recommendation [75].

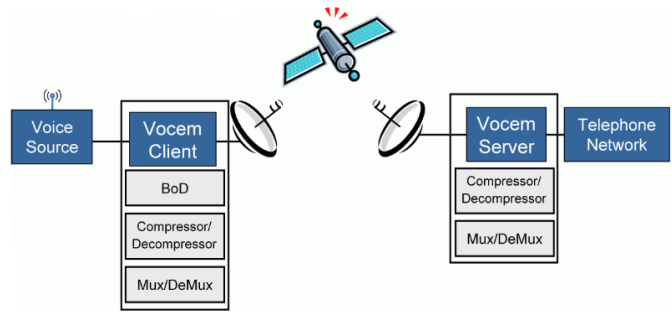

**Fig. 22** VoCeM architecture

In [79], Voice Compression and Enhanced Multiplexing (VoCeM) was proposed to optimize IuCS voice traffic from GSM and UMTS picocells and femtocells for satellite traversal. VoCeM intercepts voice traffic and then reduces its required satellite bandwidth through RTP multiplexing, header compression, and transcoding of speech frames, as illustrated in Fig. In this way, the number of calls that can be transmitted over a given satellite bandwidth is substantially increased. For example, a Broadband Global Area Network (BGAN) link has a typical capacity of 64 kbps, while an Adaptive Multi-Rate (AMR) 4.75 speech codec consumes a bandwidth of 29.2 kbps per speech stream. Hence, only two simultaneous voice streams (i.e. calls) can be served by the BGAN satellite link. However, VoCeM allows up to 11 voice streams to be transmitted simultaneously through the satellite link, thereby not only increasing service offering (i.e. more simultaneous calls supported) but also reducing the cost of satellite backhauling.

## XII. CONCLUSION

In this paper, a variety of recent topics related to small cell deployment have be presented, ranging from recent standardization efforts, SON, CA, MIMO, ICIC, energy saving techniques, ML approaches, simulation tools and deployment scenarios. While these topics are by no mean exhaustive, they collectively represent the current focuses and what are likely to come in the near future in small cell deployments.

### ACKNOWLEDGMENT

We would like to thank all workshop invited speakers and attendees. Without them the workshop would not have been possible. We also thank all previous workshops invited speakers and attendees, who have helped us improving the quality of the workshops. We also thank the EU FP7 IAPP@RANPLAN project for the financial support of the workshop.